\RequirePackage{fix-cm}
\documentclass{svjour3}                     
\smartqed  
\usepackage{graphicx}
\usepackage{amsmath,amssymb}
\usepackage{bm,color}
\usepackage[numbers]{natbib}
\begin{document}

\title{Coulomb screening effect on the Hoyle state
  energy in thermal plasmas\thanks{
This work was in part supported by JSPS KAKENHI Grants
Nos.\ 18K03635, 18H01211, 18H05406, and 19H05140.
MTY thanks the Brazilian agencies Funda\c{c}\~ao de Amparo 
\`a Pesquisa do Estado de S\~ao Paulo-FAPESP Grant 
No.\ 2019/00153-8 and Conselho Nacional de
Desenvolvimento Cient\'ifico e Tecnol\'ogico-CNPq Grant 
No.\ 303579/2019-6.
}}


\author{Lai Hnin Phyu         \and H. Moriya
  \and \\W. Horiuchi \and K. Iida \and K. Noda \and\\ M.~T. Yamashita
}


\institute{Lai Hnin Phyu \and H. Moriya \and W. Horiuchi \at
            Department of Physics, Hokkaido University, Sapporo 060-0810, Japan  \\
\email{whoriuchi@nucl.sci.hokudai.ac.jp}
\and
K. Iida \and K. Noda \at Department of Mathematics and Physics,   Kochi University, Kochi 780-8520, Japan\\
\email{iida@kochi-u.ac.jp}
\and M.~T. Yamashita \at
Instituto de F\'isica Te\'orica,
  Universidade Estadual Paulista, UNESP, 
  Rua Dr. Bento Teobaldo Ferraz, 271 - Bloco II,
  S\~ao Paulo, SP 01140-070, Brazil
}

\date{Received: date / Accepted: date}

\maketitle

\begin{abstract}
The first excited $J^{\pi}=0^+$ state of $^{12}$C, the so-called Hoyle state,
plays an essential role in a triple-$\alpha$ ($^{4}$He) reaction,
which is a main contributor to the synthesis of $^{12}$C in a burning star.
We investigate the Coulomb screening effects
on the energy shift of the Hoyle state 
in a thermal plasma environment using precise three-$\alpha$ model calculations.
The Coulomb screening effect between $\alpha$ clusters
are taken into account within the Debye-H\"uckel approximation.
To generalize our study, we utilize two standard $\alpha$-cluster models, 
which treat the Pauli principle between the $\alpha$ particles differently.
We find that the energy shifts do not depend on these models
and follow a simple estimation in the zero-size limit
of the Hoyle state when the Coulomb screening length is 
as large as a value typical of such a plasma consisting of electrons 
and $\alpha$ particles.

\end{abstract}

\section{Introduction}

The production process of the $^{12}$C element is known to be
key to understanding the nucleosynthesis in normal stars.
In a normal star, $^{12}$C is created by the triple-$\alpha$ reaction,
which involves three $^4$He ($\alpha$) nuclei via the
following two sequential steps~\cite{Salpeter52}: 
Firstly two $\alpha$ particles form a $^8$Be resonant state, and secondly
this $^8$Be fuses with another $\alpha$ particle, leading to $^{12}$C. 
To explain this process, Hoyle proposed the existence of
a resonance state of $^{12}$C with $J^{\pi} = 0^+$ at an energy just
above the three-$\alpha$ threshold, called the Hoyle state~\cite{Hoyle54},
which is essential for increasing the production rate of $^{12}$C.

Considering a realistic situation in the burning star,
one may think that
the triple-$\alpha$ process occurs only in thermal plasmas,
but this is not the case.
A cold and dense plasma environment is also suggested
in Ref.~\cite{Lewin93} in the outer layer of an X-ray bursting,
accreting neutron star
where a plasma consists of helium and electrons appears.
In this environment,
the standard Coulomb potential between approaching helium ions
is screened by surrounding degenerate electrons~\cite{Salpeter54},
which reduces the Coulomb barrier.
This Coulomb screening phenomenon can affect
the triple-$\alpha$ reaction rate as it
shifts the energy of the Hoyle state. 

The purpose of this paper is to investigate
the energy shift of the Hoyle state due to the Coulomb screening
based on precise three-$\alpha$ cluster model calculations,
mainly focusing on the model dependence of
the treatment of the $\alpha$ clusters.
The paper is organized as follows.
In Sect.~\ref{theory.sec}, we explain the theoretical models employed
in this paper. First, we briefly describe the Coulomb screening
in thermal plasmas and explain its incorporation into 
the three-$\alpha$ cluster models.
Section~\ref{results.sec} presents our results.
In Ref.~\cite{lhp20}, the three-$\alpha$ wave functions 
were obtained using a shallow potential model. In addition to them,
we show the results from the orthogonality condition model (OCM)
and compare them with the previous calculations.
Differences of these theoretical models are discussed
through a comparison of the pair density distributions
of the Hoyle state at various screening conditions.
Conclusion is drawn in Sec.~\ref{conclusion.sec}.

\vspace{-0.3cm}
\section{Coulomb screened three-$\alpha$ model}
\label{theory.sec}

To consider the Coulomb screening effect on the Hoyle state
in thermal plasmas, we apply the Debye-H\"{u}ckel approximation.
For more detail, the reader is referred to Ref.~\cite{lhp20}.
In this approximation, all ions are considered to be point charges.
This assumption leads to the screened Coulomb potential,
which is the product
of the bare Coulomb potential and $\exp(-Cr_{ij})$ with $r_{ij}$ being
the distance between the $i$th and $j$th colliding ions.
Here we  assume that the plasma consists of electrons and $\alpha$ particles,
and $C$ is the Coulomb screening factor that is the reciprocal of
the Debye screening length,
\begin{align}
  \lambda_D = \sqrt{\frac{k_B T}{4\pi e^2\left( n_e + 4 n_\alpha \right)}},
\label{Debye1}
\end{align}
where $n_e$ and $n_\alpha$ are the averaged number density
of electrons and $\alpha$ particles in the plasma with temperature $T$.
Note that charge neutrality ensures $n_e=4n_\alpha$.
The energy shift due to the Coulomb screening
$\Delta E_C$ can roughly be estimated as a difference between
the screened and bare Coulomb potential as
\begin{align}
\Delta E_C\approx\sum_{i<j}
\left\langle\frac {4e^2}{r_{ij}} \exp(- Cr_{ij})\right\rangle
-\left\langle\sum_{i<j}\frac {4e^2}{r_{ij}}\right\rangle.
\end{align}
In case that the distance between the $\alpha$ particles
is shorter than the Debye screening length,
the energy shift can be easily evaluated by using the Taylor expansion as
\begin{align}
\Delta E_C  = -\frac{12e^2}{\lambda_D} +  \mathcal{O}\left( \langle r_{ij}\rangle \right),
\label{DEC1}
\end{align}
which suggests that the $Q$ value shift $\Delta Q$ by the 
Coulomb screening effect is well approximated by $12e^2/\lambda_D$
in both the weak screening limit and the zero-size limit of the Hoyle state.

To discuss the energy shift of the Hoyle state,
we incorporate the Coulomb screening effect
into three-$\alpha$ Hamiltonian.
The Hamiltonian for the three-$\alpha$ system is generally written as
\begin{align} 
  H = \sum_{i=1}^{3} T_i   -T_{cm}  + \sum_{i<j}
  \left[ V_{ij}^{2\alpha}  + V_{ij}^{\rm Coul}(C) \right]
+  V_{123}^{3\alpha}, 
\label{hamiltonian}
\end{align}
where $T_i$ is the kinetic energy operator of the $i$th $\alpha$ particle,
$T_{\rm cm}$ is the center-of-mass kinetic energy, $V^{2\alpha}_{ij}$
and $V^{3\alpha}_{123}$ are respectively two- and three-$\alpha$ potentials,
and $V^{\rm Coul}_{ij}$ is the screened Coulomb potential, which we take
\begin{align}
  V^{\rm Coul}_{ij}(C)=\frac{4e^2}{r_{ij}}\exp(- Cr_{ij}).
\end{align}
Note that
$C=0$ means no Coulomb screening corresponding
to the bare Coulomb interaction, while $C\to \infty$ means
full screening corresponding to no Coulomb interaction.
In reality, an $\alpha$ particle is not a point charge.
Such an finite size effect is taken into account
by assuming the Gaussian-type charge form factor~\cite{lhp20}.

To investigate the model dependence,
we examine two types of potential models, namely,
shallow and deep potential models for the inter-cluster potentials,
$V^{2\alpha}_{ij}$ and $V^{3\alpha}_{123}$.
For the shallow potential model,
we employ two different $V^{3\alpha}_{123}$:
One includes only attraction~\cite{AB,Fedorov96,Suno15} (Set 1), while
the other does an attraction with a short-range repulsion~\cite{lhp20} (Set 2).
In addition to these two models, here we employ
another standard $\alpha$ cluster model,
the orthogonality condition model (OCM~\cite{OCM}),
which treats the Pauli principle between $\alpha$ particles differently.
It is known that the shallow potential model and OCM
give different results in observables~\cite{Pinilla11,Arai18}.
In the OCM case, we adopt the same Hamiltonian
used in Refs.~\cite{kk05,kk07}.
For the OCM calculation, the potential is deep and generates
some redundant forbidden states. To impose the orthogonality condition
on the Pauli forbidden states, we practically introduce
a pseudo potential that projects the relative wave function
on the Pauli forbidden states~\cite{Kukulin78}.
By multiplying the projection operator by 
a large prefactor, e.g., $10^5$ MeV in the present case,
the forbidden states
are ruled out variationally from the total three-$\alpha$ wave function.

Once the Hamiltonian is set, we solve the three-$\alpha$
Schr\"odinger equation and obtain precise three-$\alpha$ wave functions
using the stochastic variational method
with a fully symmetrized correlated Gaussian basis~\cite{Varga,SVM}.
More explicitly, the wave function of
the three-$\alpha$ system is expressed as
\begin{align}
 \Psi =  \sum_{k=1}^{K} c_k  \mathcal{S}  G(k,\bm{x}_1,\bm{x}_2)
 \end{align}
with the symmetrizer $\mathcal{S}$ and the correlated Gaussian basis,
 \begin{align}
 G(k,\bm{x}_1,\bm{x}_2)&=\exp\left(-\frac{1}{2}A_{11}^kx_1^2-\frac{1}{2}A_{22}^kx_2^2
 -A_{12}^k\bm{x}_1\cdot\bm{x}_2\right),
 \label{gaussian}
 \end{align}
 where $A_{11}^k$, $A_{22}^k$, and $A_{12}^k$ are the variational parameters
 for the $k$th basis.
 Note that the two Jacobi coordinates $\bm{x}_1=\bm{r}_1-\bm{r}_2$
 and $\bm{x}_2=(\bm{r}_1+\bm{r}_2)/2-\bm{r}_3$
 are  explicitly correlated via $A_{12}$,
 describing a complicated correlation
 among $\alpha$ particles.
 One notable advantage of the use of the correlated Gaussian basis
 is that its functional form does not change under any coordinate
 transformation, which greatly simplifies the manipulation
 of the symmetrization of the basis.
 See Refs.~\cite{Suzuki08,Mitroy13,Suzuki17} for many
 successful applications of the correlated Gaussian basis.
 Full details for the optimizations of a number of those variational
 parameters are given in Ref.~\cite{lhp20}.

\vspace{-0.3cm}
\section{Results and discussions}
\label{results.sec}

\begin{figure}
  \begin{center}
    \includegraphics[width=9cm]{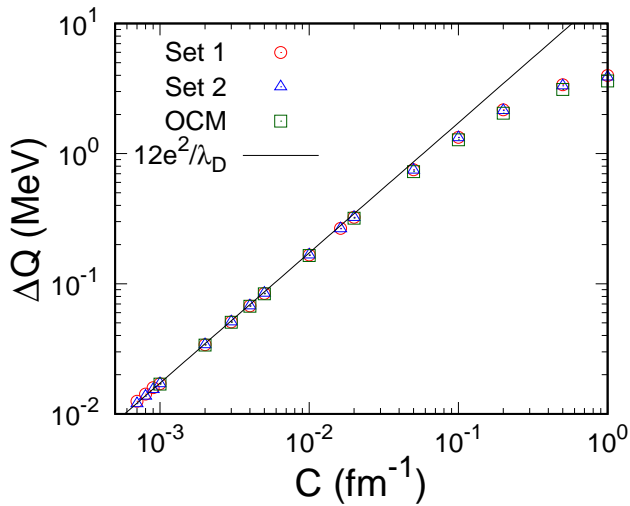}
  \end{center}
  \caption{Screening-induced $Q$ value shifts of the Hoyle state
  calculated as a function of Coulomb screening factor $C$.}
  \label{Qshift1.fig}      
\end{figure}

Figure~\ref{Qshift1.fig} shows the $Q$ value shifts
$\Delta Q=-\Delta E_C$
of the Hoyle state as a function of the Coulomb screening factor $C$.
Although the Hamiltonian is different among Set 1, Set 2, and OCM,
we find the results are almost identical.
The $\Delta Q$ values coincide with the simple estimation
of the $Q$ value shift, $12e^2C=12e^2/\lambda_D$,
for small $C\lesssim 0.2$ fm$^{-1}$.
It should be noted that
in the plasma environment considered in this paper,
the temperature $T$ is $\approx 10^8$ K and
the density $\rho$ is $\approx 10^3$--$10^6$ g cm$^{-3}$.
The resulting Debye screening length is 
of the order of 10$^{3}$--10$^{4}$ fm
($C=10^{-3}$--$10^{-4}$ fm$^{-1}$),
where the simple estimation of Eq.~(\ref{DEC1}) is valid.
This can simply be understood from the fact that, for instance,
$\lambda_D=100$ fm is much larger than the root-mean-square pair distance
$d=\sqrt{\left<\Psi\right|\bm{x}_1^2\left|\Psi\right>}$
of the screened Hoyle states, which are obtained for different models
as $\approx 6$--8 fm.
At $C \gtrsim 0.1$ fm$^{-1}$, where the Coulomb screening length
becomes $\lesssim 10$ fm, the Coulomb shift deviates
from the zero-size limit as the screening length is comparable
to the $d$ value of the screened Hoyle state, e.g.,
$d=6$--7 fm at $C=0.1$ fm$^{-1}$.

\begin{figure}[ht]
  \includegraphics[width=0.5\hsize]{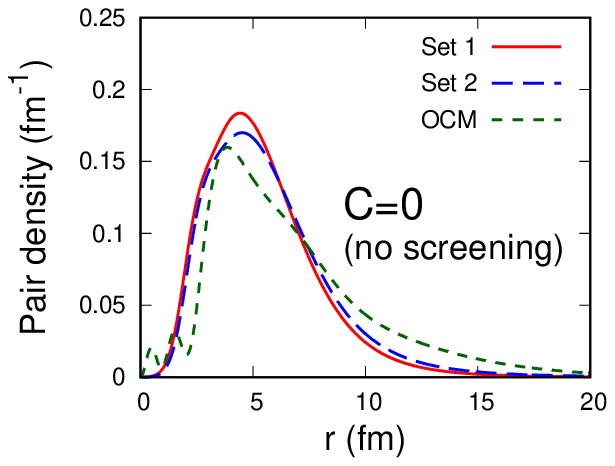}
  \includegraphics[width=0.5\hsize]{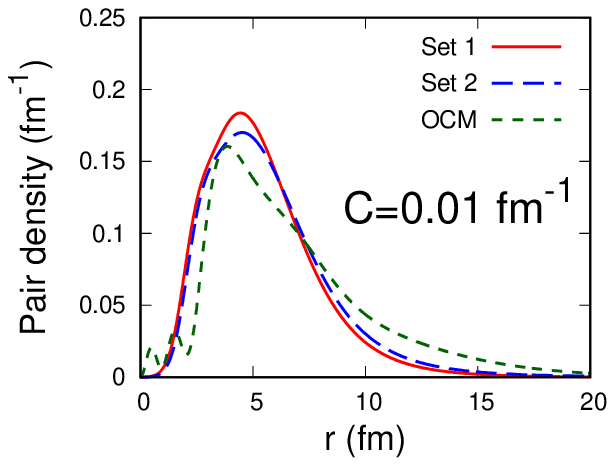}
  \includegraphics[width=0.5\hsize]{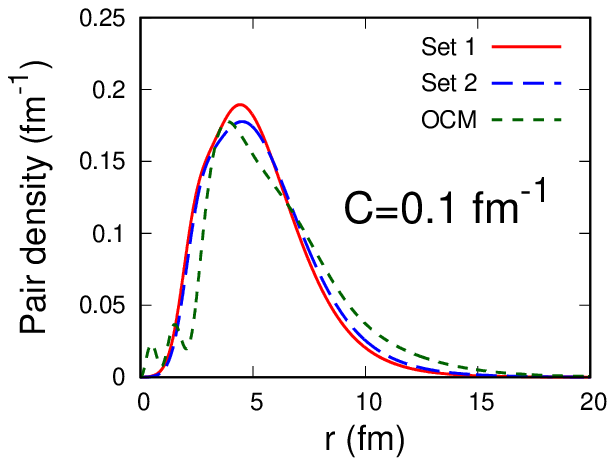}
  \includegraphics[width=0.5\hsize]{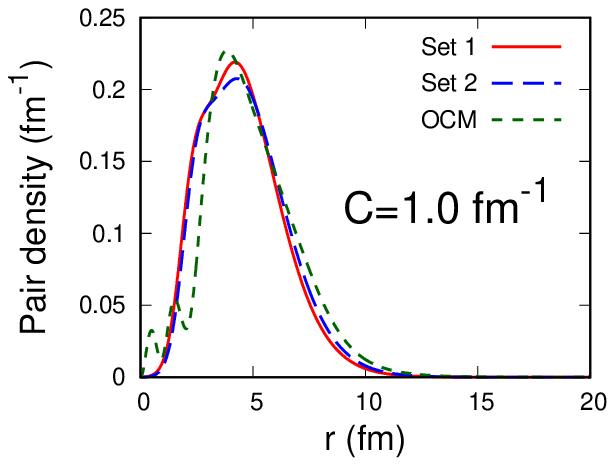}
  \caption{Pair densities of the Hoyle state without ($C=0$)
   and with the Coulomb screening for $C=0.01, 0.1, 1.0$ fm$^{-1}$.}
\label{density.fig}    
\end{figure}

Although there is no appreciable model dependence in the $Q$ value shifts,
it is interesting to see how the internal structure of the wave function
depends on the different treatments of the Pauli-forbidden
states~\cite{Arai18}.
For this purpose, we calculate
the pair density of the three-$\alpha$ system defined as
\begin{align}
  \rho_{\rm pair}=\left<\Psi\right|\delta(|\bm{x}_1|-r)\left|\Psi\right>,
\end{align}
which satisfies $\int_0^\infty dr \rho_{\rm pair}(r)=1$ and
$\int_0^\infty dr r^2\rho_{\rm pair}(r)=d^2$.
Figure~\ref{density.fig} plots the pair densities of
the Hoyle state in vacuum ($C=0$, no screening) and
the screened Hoyle state with $C = 0.01, 0.1$, and 1.0 fm$^{-1}$. 
For all the $C$ values, we see the results with Sets 1 and 2 potentials
are similar, while OCM equally shows some nodal behavior at short distances
at $r\lesssim 3$ fm due to the orthogonality condition
imposed on the total three-$\alpha$ wave function.
This is expected because the Coulomb screening only affects
the outer region of the wave function.
At small $C=0.01$ fm$^{-1}$, 
the screening length is 100 fm, which is so large that
the density tails are not much affected.
At $C=0.1$ fm$^{-1}$, all states are bound
below the energy of the ground state of $^{8}$Be~\cite{lhp20}.
The screening length $\lambda_D=10$ fm is slightly larger
than the $d$ value of the screened Hoyle state.
Accordingly, the pair density at $r\gtrsim 10$ fm is reduced.
At $C=1$ fm$^{-1}$, the screening length is 1 fm,
which is so small that the Coulomb interaction is screened
almost completely.
The behavior of the tail structure becomes essentially the same
as all states are deeply bound.

\vspace{-0.3cm}
\section{Conclusion and future perspectives}
\label{conclusion.sec}

In this paper, we have studied the Coulomb screening effect
on the Hoyle states in thermal plasmas
using precise three-$\alpha$ model calculations.
To see the effect of the Pauli principle on
the internal structure,
we show the results obtained by the orthogonality condition model (OCM),
which is one of the standard three-$\alpha$ cluster models.
Basically, the same conclusion as that given in Ref.~\cite{lhp20} is obtained,
although we see some differences of the internal structure
in the pair density distributions between the shallow potential model and OCM.
As long as the Coulomb screening is small as in normal stars, 
the energy shift is universal, which suggests that it is insensitive to the 
internal structure of the three-$\alpha$ wave function.

To extend this work, it is interesting to explore
the existence and stability of three-$\alpha$ systems
in various astrophysical environments.
Given that the effective mass of the $\alpha$ particle
could be changed in cold neutron matter~\cite{Nakano20},
the structure of the Hoyle state may also be affected
depending on the density and temperature of the neutron matter.
The work along this direction is in progress~\cite{Moriya21}.

\end{document}